\documentclass[twocolumn,twoside,letterpaper,11pt]{article}
\usepackage{graphicx,xrrc,natbib}
\usepackage{times,latexsym}

\begin{document}

\title{PARSEC-SCALE X-RAY FLOWS IN HIGH-MASS STAR-FORMING REGIONS}



\author{L. Townsley} 
\institute{Penn State University, Department of Astronomy \& Astrophysics} 
\address{525 Davey Laboratory, University Park, PA 16802, U.S.A.} 
\email{townsley@astro.psu.edu}
\author{E. Feigelson, T. Montmerle, P. Broos}
\email{edf@astro.psu.edu, montmerle@obs.ujf-grenoble.fr, patb@astro.psu.edu}
\author{Y.-H. Chu, G. Garmire, K. Getman}
\email{chu@astro.uiuc.edu, ggarmire@astro.psu.edu, gkosta@astro.psu.edu}


\maketitle

\abstract{ We present {\em Chandra}/ACIS images of several high-mass
star-forming regions.  The massive stellar clusters powering these HII
regions are resolved at the arcsecond level into hundreds of stellar
sources, similar to those seen in closer young stellar clusters.
However, we also detect diffuse X-ray emission on parsec scales that is
spatially and spectrally distinct from the point source population.
For nearby regions (e.g.\ M17 and Rosette) the emission is soft, with
plasma temperatures less than 10 million degrees, in contrast to what
is seen in more distant complexes (e.g.\ RCW49, W51).

This extended emission most likely arises from the fast O-star winds
thermalized either by wind-wind collisions or by a termination shock
against the surrounding media.  We have established that only a small portion
of the wind energy and mass appears in the observed diffuse X-ray
plasma; in the blister HII regions, we suspect that most of it flows
without cooling into the low-density interstellar medium through
blow-outs or fissures in the surrounding neutral material.  These data
provide compelling observational evidence that strong wind shocks are
present in HII regions. }

\section{Introduction}

Most stars in the Galactic disk form in high-mass star-forming regions
(HMSFRs) where rich clusters containing thousands of stars are produced
in massive molecular cores.  X-ray surveys of HMSFRs are important
because they readily discriminate young stars from unrelated objects
that often contaminate $JHK$ images of such fields, especially for
those young stars no longer surrounded by a dusty circumstellar disk.
Radio observations opened up studies of embedded star formation,
revealing ionized gas despite large absorbing columns.

Plasmas at keV and even MeV energies are prevalent in young stellar
systems, where we find wind-generated X-rays in OB stars, magnetic
flare-generated X-rays in pre-main sequence lower-mass stars, and
shocks from OB winds and supernovae interacting with the ambient
medium.  These hot plasmas are critical drivers of galactic evolution,
injecting energy and enriched material into the interstellar medium
(ISM) through wind bubbles, supernova remnants (SNRs), superbubbles,
and chimneys to the Galactic halo.  HMSFRs are thus likely to exhibit a
complicated mixture of pointlike and extended structures which are easily
confused by low-resolution X-ray telescopes.  Before the launch of the
Chandra X-ray Observatory \citep[{\em Chandra},][]{Weisskopf02}, the
relative X-ray contributions of high-mass and low-mass stars, OB winds,
and SNR shocks in these regions was largely unknown.  {\em Chandra},
with its Advanced CCD Imaging Spectrometer (ACIS) camera
\citep{Garmire03}, starts a new chapter in star formation studies,
finally giving us the sensitivity, spatial resolution, and broad
bandpass to detect diffuse X-ray emission and to separate it from the
hundreds of X-ray-emitting stars in HMSFRs.

The following section provides a brief overview of wind-blown bubbles
and their effect on the classical definition of an HII region.  We then
describe our own {\em Chandra} observations of several HMSFRs and give some
spectral characterizations of the diffuse X-rays seen there; we start
with a review of M17, where the case for diffuse X-ray emission is
clear.  We continue with some preliminary images and spectra of RCW49,
a southern HII region powered by a massive, condensed stellar cluster
whose morphology in the radio and IR has been significantly altered by
the presence of two Wolf-Rayet stars.  We complete our simple survey
with our {\em Chandra} data on W51A, a very massive, young star-forming
complex in the Sagittarius Arm containing over 100 O stars and dozens
of radio HII regions; we see diffuse and/or point-like X-ray sources
associated with a large fraction of these radio HII regions.  We then
summarize the properties of diffuse X-ray emission in a variety of
HMSFRs studied with {\em Chandra} and {\em XMM-Newton} and comment on future prospects for
this field.

\section{ISM Bubbles and Hollow HII Regions}

The physics of the radiative ionization of HII regions worked out
during the 1930-50s by Str{\" o}mgren and others omitted the role of OB
winds, which were not discovered until the 1960s \citep{Morton67}.  The
winds play a small role in the overall energetics of HII regions but
they dominate the momentum and dynamics of the nebula.  The earliest O
stars are subject to a large mass-loss ($\dot{M} > 10^{-6} M_\odot$/yr,
$v_w \simeq 1000$--$2500$~km/s), converting several percent of the
radiative luminosity into wind mechanical luminosity with $\frac{1}{2}
\dot{M} v_w^2 \sim 10^{36-37}$~ergs/s.  A naive estimate shows that the
thermalization of the wind should yield post-shock energies of several
keV, with luminosities greatly exceeding that produced close to the
star.  In the seminal model of \citet{Weaver77}, an O star will create
a ``wind-swept bubble'' with concentric zones:  a freely expanding wind,
a wind termination shock followed by an X-ray emitting zone, the
standard $T=10^4$~K HII region, the ionization front, and the interface
with the cold interstellar environment.  \citet{Capriotti01} review the
possible relationships between ionizing radiation and winds in HII
regions.  There was thus considerable reason to expect diffuse X-ray
emission within HII regions excited by early O stars.

While a considerable literature exists on both the theory and
observation of these X-ray emitting wind shocks close to the star,
where $L_x \sim 10^{-7} L_{bol} \sim 10^{32-33}$~ergs/s, the X-ray
signature of the wind shocks on parsec scales where the $10^4$~K HII
region appears had not been clearly seen before {\em Chandra}.  A closer look
at the interaction between the winds and the surrounding HII region
shows that the problem is complex and still unresolved.  One
long-recognized problem was that the Weaver et al.\ model predicts a
much larger size for the cavity than is observed.  A recent kinematic
analysis of the Rosette Nebula, for example, yields a characteristic
age of a few $10^4$~yrs for the Rosette cavity, 100 times smaller
than the age of the exciting stars \citep{Bruhweiler01}.  Perhaps the
wind energy is dissipated in a turbulent mixing layer between an
ionization-bounded HII layer and a hot, shocked stellar wind
\citep{Kahn90}, or the OB winds are ``mass loaded,'' entraining
interstellar gas and thereby weakening the terminal shock
\citep[e.g.][]{Pittard01}.  Another complication is that the winds from
several OB stars may collide and shock before they hit the ambient
medium \citep{Canto00}.  These issues are also of great importance to
our understanding of superbubbles and the dynamics of the interstellar
medium on a Galactic scale.

We recently reported the {\em Chandra} discovery of diffuse X-rays in
the Rosette Nebula and M17 \citep[][hereafter Townsley03]{Townsley03}.
Our study not only produced images of diffuse emission with the stellar
sources removed, but made a physical characterization of the
X-ray-emitting gas.  For both Rosette and M17, we find a center-filled
morphology with $M \sim 0.1$~M$_\odot$ of $T = 7$--9~MK plasma with
density 0.1--0.3~cm$^{-3}$ spread over several cubic parsecs.  No
heating, cooling, or shocks are seen, although our sensitivity to
substructures is limited.  In Rosette, the X-ray plasma fills the
center of an annulus of $T = 10^4$~K plasma, as predicted by Weaver et
al.  But in both regions the X-ray plasma represents only 4000 years of
wind production and 20\% of the wind kinetic energy, implying that most
of the wind material and energy flows unimpeded away from the molecular
cloud.

To gain more insight into the fate of OB winds in HII regions, we are
in the process of obtaining and analyzing more {\em Chandra} observations of
high-mass star-forming regions, with a range of ages, OB membership,
and environments.  As an introduction to some of our newer {\em Chandra}
studies (on RCW49 and W51A), we begin with a short review of the
Chandra results on M17 from Townsley03.

\section{M17, The Omega Nebula}

Perhaps the clearest example of diffuse X-ray emission in HMSFRs is our
{\em Chandra} observation of M17, a bright blown-out blister HII region
on the edge of a massive molecular cloud at a distance of 1.6~kpc,
where $10^{\prime} \sim 4.7$~pc.  We summarize the Townsley03 results
here because the edge-on geometry of M17 makes it clear that the
diffuse soft X-rays are spatially distinct from the majority of the
stellar X-ray sources; {\em Chandra} allows us to remove the point source
contribution with very high precision, enabling us to characterize the
hot, probably stellar wind-generated plasma flowing out of the HII
region.  Thus M17 is a straightforward case study that will help to
explain the more complicated fields RCW49 and W51A.

\begin{figure}
\centering
\includegraphics[width=\columnwidth]{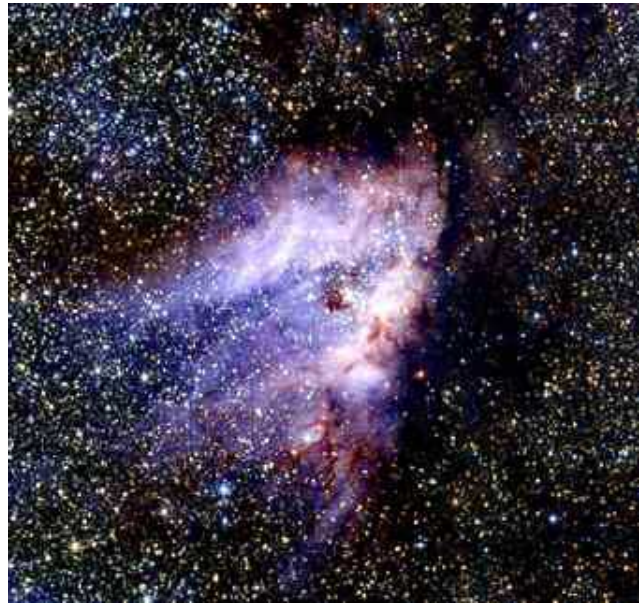} 
\caption{\small $\sim 15^{\prime} \times 15^{\prime}$ 2MASS Atlas image of M17, 
centered on the concentration of O4/O5 stars that we call the Ring of Fire.}
\label{fig:m172mass}
\end{figure}

M17 is a strong thermal radio source, showing very high ionization.  It
has been called an edge-on version of the Orion K-L region; the
expansion of the blister HII region is triggering star formation in its
associated giant molecular cloud (GMC), which contains an ultracompact HII
region, water masers, and the massive, dense core M17SW.  M17 has 100
stars earlier than B9 (Orion has 8), with 14 O stars.  Several of these
O4/O5 stars are concentrated at the core of the stellar cluster and
form what we call the Ring of Fire, $\sim 1^{\prime}$ in diameter.  The
age of the complex is estimated at $\sim 1$~Myr \citep{Hanson97}.  The
ionization front of the HII region encounters the M17 GMC
along two photodissociation regions, called the northern and southern
bars.  These are clearly seen in radio continuum data and in the 2MASS
Atlas image (Figure~\ref{fig:m172mass}), as is the Ring of Fire.

\begin{figure}
\centering
\includegraphics[width=\columnwidth]{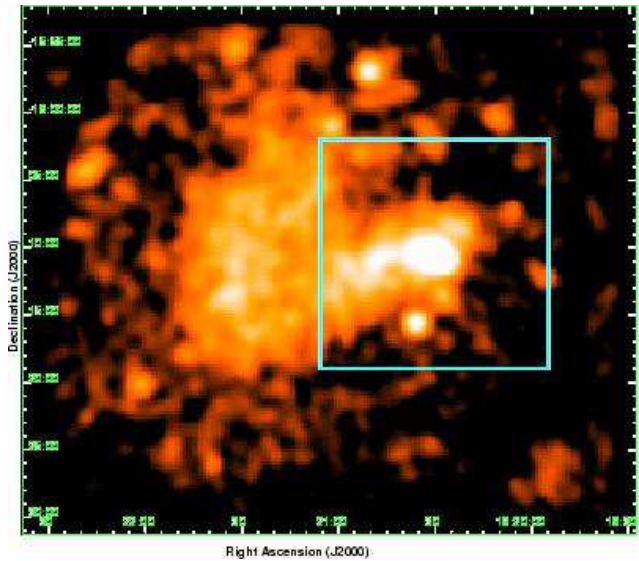}
\caption{\small Smoothed {\em ROSAT} PSPC (soft X-ray) image of M17,
$\sim 39^{\prime} \times 42^{\prime}$, with the outline of the ACIS-I
pointing overlaid in blue.}
\label{fig:m17rosat}
\end{figure}

M17 was observed in 1993 with {\em ROSAT}, but the data were only
recently published \citep{Dunne03}.  A smoothed image of the full-band
(0.1--2.4 keV) {\em ROSAT} data is shown in Figure~\ref{fig:m17rosat},
with the smaller {\em Chandra}/ACIS field of view outlined in blue.
Diffuse X-rays appear to flow away from the Ring of Fire, filling a
$\sim 20^{\prime} \times 30^{\prime}$ region eastward of the HII
region.  {\em ROSAT} detected only 5 point sources associated with the
M17 cluster; without higher spatial resolution, it is impossible to
separate the stellar X-ray emission from diffuse emission generated by
stellar wind collisions.
  
\begin{figure}
\centering
\includegraphics[width=\columnwidth]{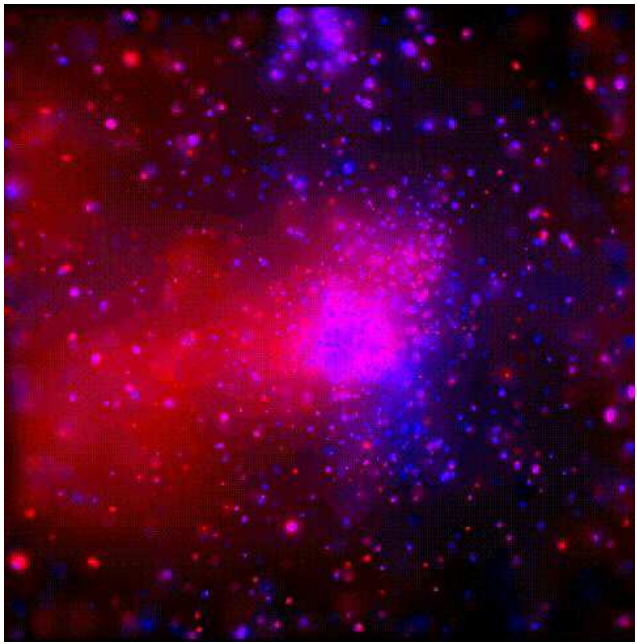}
\caption{\small The 40-ksec ACIS-I observation of M17 ($17^{\prime}
\times 17^{\prime}$), smoothed with the CIAO tool {\it csmooth}.  Red
intensity is scaled to the soft (0.5--2~keV) emission and blue
intensity is scaled to the hard (2--8~keV) emission.}
\label{fig:m17acis}
\end{figure}

\begin{figure}[htb]
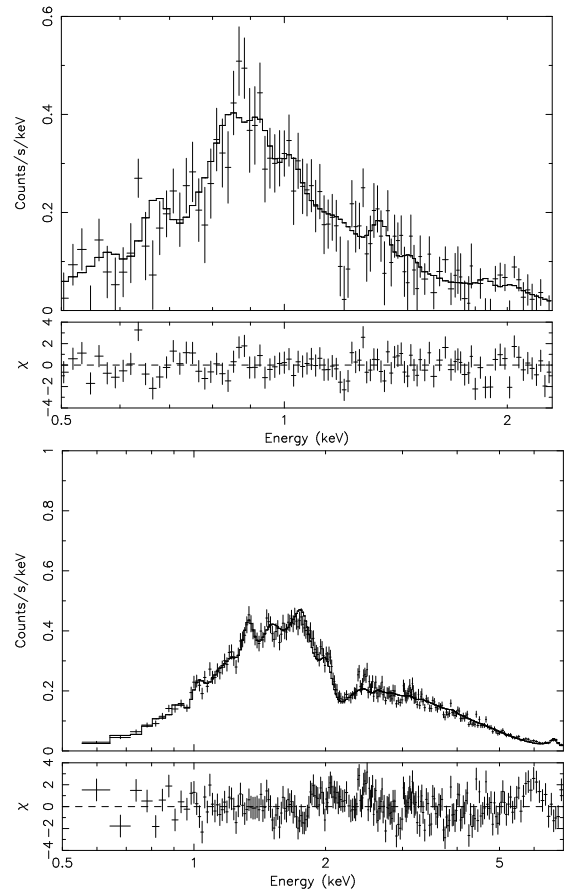

\centerline  \mbox{
\includegraphics[width=0.7\columnwidth, angle=-90]{m17_diffuse_spec_plain.ps}
\includegraphics[width=0.7\columnwidth, angle=-90]{m17_pt_spec_plain.ps} }
\caption{\small Top panels:  M17 diffuse emission spectrum and fit residuals.  Bottom panels: M17 composite point source spectrum and fit residuals.
Note the different scales and energy ranges in these plots.  
Comparison of the spectral shapes shows that the diffuse emission is not
likely to be composed primarily of unresolved point sources.}
\label{fig:m17spectrum}
\end{figure}

The {\em Chandra} observation (Figure~\ref{fig:m17acis}) illustrates
the significance of this problem.  This 40-ksec ACIS observation
reveals over 900 point sources in a $\sim 17^{\prime} \times
17^{\prime}$ field centered on the grouping of early O stars that we
call the Ring of Fire.  This point source emission dominates the
observation, with three times the X-ray luminosity of the diffuse
emission.  The diffuse emission is spectrally distinct from the
integrated point source emission (Figure~\ref{fig:m17spectrum}) and is
spatially concentrated eastward of the Ring of Fire, centrally filling
the region delineated by the two bars seen in the 2MASS and radio
data.

The M17 diffuse emission is adequately fit by a two-temperature thermal
plasma model and a single absorbing column, with $kT=0.13$ and 0.6~keV,
$N_H=4 \times 10^{21}$~cm$^{-2}$, and a total intrinsic X-ray
luminosity (corrected for absorption) of $L_{{\rm X,diffuse}}=3.4
\times 10^{33}$~ergs/s.  The composite point source spectrum, on the
other hand, requires a much harder thermal plasma ($kT=3$~keV) and
higher absorption ($N_H=17 \times 10^{21}$~cm$^{-2}$) and is much
brighter ($L_{{\rm X, pt srcs}}=10.2 \times 10^{33}$~ergs/s); details
are given in Townsley03.

The {\em Chandra} data confirm and clarify the impression from the {\em
ROSAT} data that there is a hot plasma flowing out of the HII region,
probably generated by wind-wind collisions between the massive stars in
the Ring of Fire.  Notably, we do not measure any significant
temperature change in the plasma with position away from the O stars;
this implies that the hot gas flows into the ambient ISM without
substantial cooling.  An additional hard plasma component may be
present, spatially coincident with the highest concentration of X-ray
point sources near the cluster center, but this may be a population of
embedded, unresolved point sources.  As will be demonstrated below,
such a hard diffuse plasma component may be present in dense, massive, young
clusters; if so, it indicates that an additional, more efficient
process is at work to convert the wind kinetic energy into the thermal
energy of the plasma.

\pagebreak

\section{RCW49 and Westerlund 2}

RCW 49 is a prominent giant H II region, covering $90^{\prime} \times
70^{\prime}$ in the southern sky.  Both point source and extended
X-rays were detected by {\em ROSAT} and {\em Einstein}.  Neither
previous observation, though, had the spatial resolution to separate
truly diffuse emission from the point source population.

RCW 49 is powered by Westerlund 2 (W2), an OB association containing at
least a dozen OB stars: the earliest is an O6 star, there are 5 O7
stars, and two Wolf-Rayet (W-R) stars \citep{Moffat91}.  The presence
of these massive stars implies a cluster age of 2-3 Myr
\citep{Piatti98}.  The distance to RCW 49 is uncertain, with estimates
in the range 1.9--7.9~kpc; here we adopt $D = 2.3$~kpc (where
$10^{\prime} \sim 6.7$~pc).

\begin{figure}
\centering
\includegraphics[width=\columnwidth]{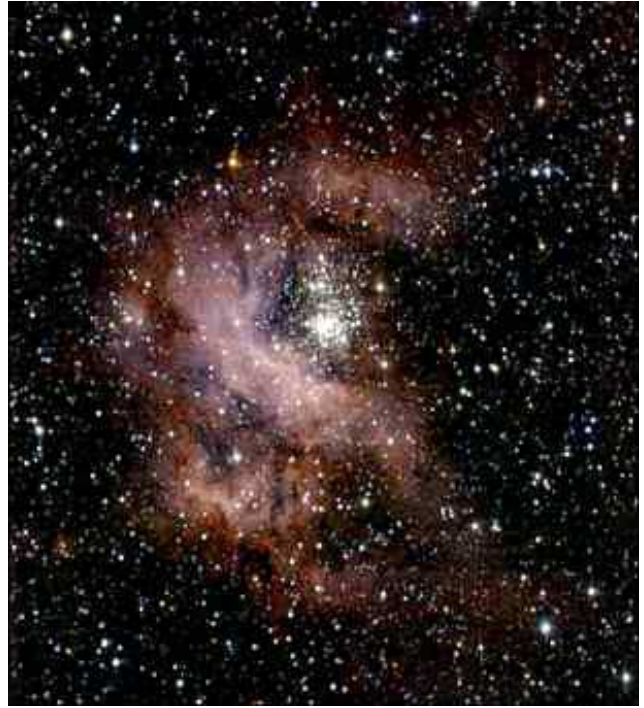}
\caption{\small $\sim 16^{\prime} \times 16^{\prime}$ 2MASS Atlas image
of RCW49 centered on its ionizing cluster Westerlund 2.}
\label{fig:rcw492mass}
\end{figure}

Radio continuum observations \citep{Whiteoak97} revealed the presence
of two wind-blown shells at the center of RCW 49.  One of these is
closed and is most likely blown by the W-R star WR20b.  The other shell
is centered on the OB association, but there is another W-R star
(WR20a) in this region as well; because of its strong wind, it probably
governs the dynamics of the shell.  This shell shows a large blister
extending to the west in the radio data.  Figure~\ref{fig:rcw492mass}
shows the 2MASS Atlas image of RCW49.  The closed wind-blown shell
associated with WR20b is clearly seen, southeast of the stellar
cluster.  The shell associated with W2 and WR20a appears closed on
three sides, opening to the west.

\begin{figure}
\centering
\includegraphics[width=\columnwidth]{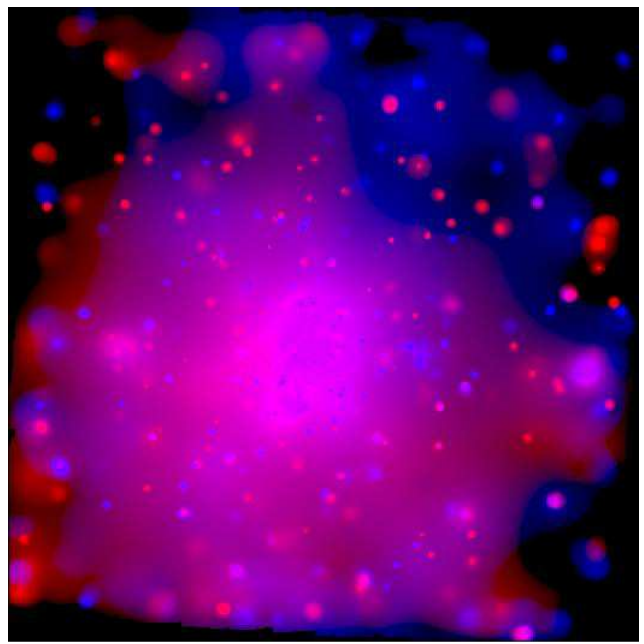}
\caption{\small The 36-ksec ACIS-I observation of RCW49
($17^{\prime} \times 17^{\prime}$), smoothed with {\it
csmooth}.  Colors: red = (0.5--2~keV), blue = (2--8~keV). }  
\label{fig:rcw49acis}
\end{figure}

ROSAT found 3 point sources associated with W2, plus diffuse emission
pervading RCW49 \citep{Belloni94}.  A smoothed image of our 36-ksec
{\em Chandra}/ACIS-I observation of RCW49 is shown in
Figure~\ref{fig:rcw49acis}.  Over 500 point sources are detected,
including both W-R stars.  Over 100 of those point sources are
spatially coincident with the W2 cluster.  Even though {\em Chandra} resolves
much of the {\em ROSAT} extended emission into point sources, a diffuse
component remains, centered on the ionizing cluster W2.  This diffuse
emission is not obviously associated with the W-R stars.

\begin{figure}
\centering
\includegraphics[width=\columnwidth]{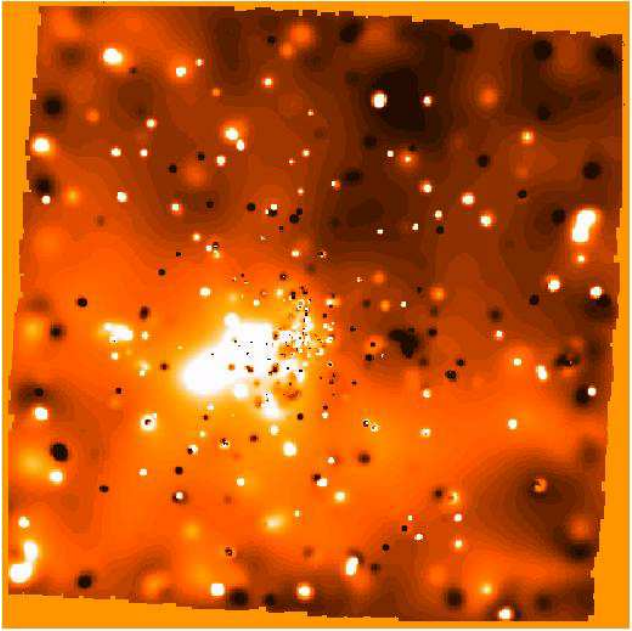}
\caption{\small A hardness-ratio image of RCW49 (again $17^{\prime}
\times 17^{\prime}$):  lighter regions are dominated by soft X-rays,
darker regions are dominated by hard X-rays.}
\label{fig:rcw49hardness}
\end{figure}

To give a clearer indication of the spectral variation across the ACIS
field, we have generated an image of the spectral hardness
(Figure~\ref{fig:rcw49hardness}).  This image shows a pronounced area
of soft diffuse emission extending eastward from W2 $\sim 2^{\prime}$,
a large expanse of slightly harder diffuse X-rays covering much of the
lower half of the image, and patches of harder diffuse emission,
possibly associated with clumps of point sources, to the north and
west.  The mixing of soft and hard point sources across the field
indicates that sources lie both in front of and behind RCW49's GMC.

\begin{figure}
\centering
\includegraphics[width=\columnwidth]{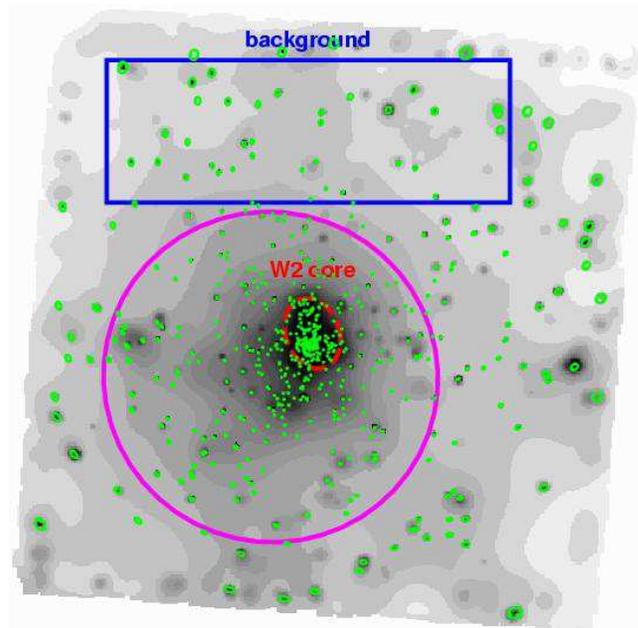}
\caption{\small  The full-band (0.5--8 keV) ACIS-I image of RCW49 with
detected point sources outlined in green.  Regions used for obtaining
the spectrum of the diffuse emission are described in the text.}
\label{fig:rcw49fullband}
\end{figure}

The full-band smoothed ACIS-I image of RCW49 with its $>500$ detected
point sources outlined in green is shown in
Figure~\ref{fig:rcw49fullband}; also shown are regions used for
spectral fitting of the diffuse emission.  The region used to make the
diffuse spectrum is the magenta circle (with point source regions and
the area outlined in red excluded); the background spectrum was
obtained from the blue box.  The background almost certainly contains
diffuse emission as well as true background; we chose to suffer the
loss of diffuse photons in our background-subtracted spectrum because
the spectral background on the Galactic plane is complex and not
well-represented by generic backgrounds taken from other parts of the
sky.  The region outlined in red was excluded from our sample of the
diffuse X-ray emission because of its high point source density.

\begin{figure}
\centering
\includegraphics[width=0.7\columnwidth,angle=-90]{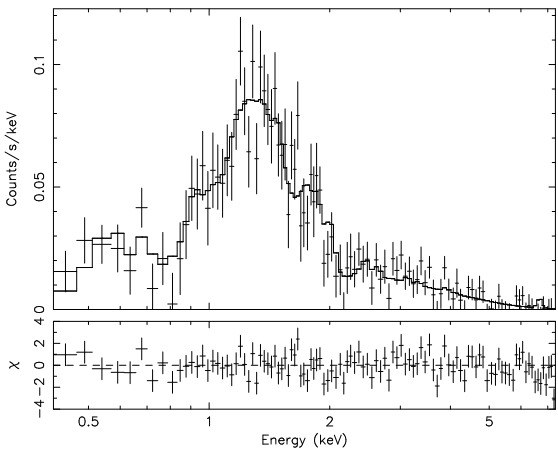}
\caption{\small RCW49 diffuse emission:  a transition example, showing
both the soft diffuse emission seen in Rosette and M17 and a harder
component, similar to that seen in more distant HMSFRs. }
\label{fig:rcw49spectrum}
\end{figure}

The background-subtracted spectrum and a model fit are shown in
Figure~\ref{fig:rcw49spectrum}.  The model consists of three thermal
plasma components and two absorbing columns; the very soft component
($kT = 0.1$~keV) is absorbed by $N_H = 4 \times 10^{21}$~cm$^{-2}$,
while the two harder components ($kT = 0.8, 3.1$~keV) both require an
absorbing column of $N_H = 12 \times 10^{21}$~cm$^{-2}$.  The hardest
thermal component is not well-constrained by these data; a thermal
plasma ($kT \sim 3$~keV) or a power law ($\Gamma = 2.3$) both give
acceptable fits to the high-energy tail in the spectrum.  In order to
achieve an acceptable fit, we also had to add a gaussian component at
1.2~keV.  The intrinsic luminosity (absorption-corrected, 0.5--8~keV)
of the $kT \sim 0.1$~keV component is $L_{{\rm X,diffuse}}=1.3 \times
10^{33}$~ergs/s; for the $kT \sim 0.8$~keV component we get $L_{{\rm
X,diffuse}}=0.8 \times 10^{33}$~ergs/s, while that of the $kT \sim
3$~keV component is $L_{{\rm X,diffuse}}=0.9 \times 10^{33}$~ergs/s.

From Townsley03, we expect the intrinsic X-ray luminosity of diffuse
emission to be roughly $L_{{\rm X,diffuse}} = 2-5 \times
10^{32}$~ergs/s per early O star for HMSFRs similar to Rosette and
M17.  W2 has 8 stars (including the two W-R stars) earlier than O7, so
if our rough scaling holds we would expect the diffuse emission in RCW
49 to have an intrinsic luminosity of $L_{{\rm X,diffuse}} = 1.6-4
\times 10^{33}$~ergs/s.  This is consistent with what we measure from
the ACIS spectrum.

Despite the simplicity of this analysis, it does show that RCW49 is a
very useful ``transition'' target, exhibiting in its diffuse X-rays
both the soft spectral components seen in Rosette and M17 and a hard
tail.  The hard diffuse component may be related to the high
concentration of stars in W2 -- other dense clusters show a similar
hard thermal plasma \citep[e.g.\ NGC~3603,][]{Moffat02}.
Alternatively, if this hard tail is non-thermal, it may be the spectral
signature of an embedded SNR, as suggested by \citet{Wolk02} to explain
the {\em Chandra} spectrum of diffuse X-rays in RCW38.

\section{W51A}

\begin{figure}
\centering
\includegraphics[width=\columnwidth]{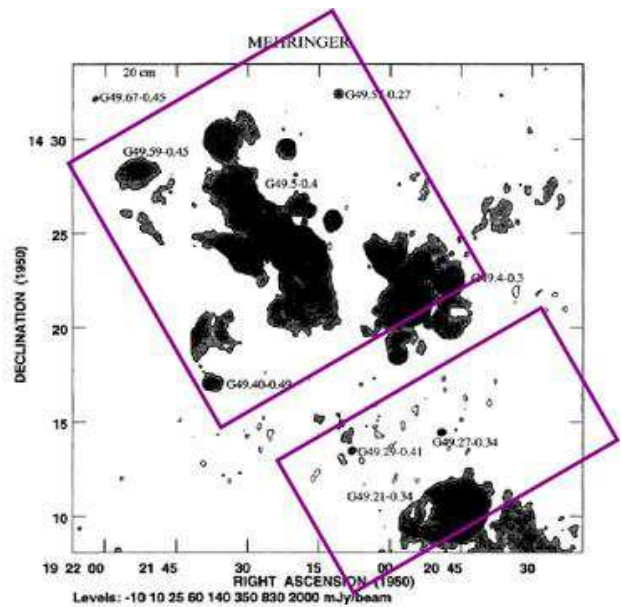}
\caption{\small A radio continuum image of W51 from \citet{Mehringer94}
showing many HII regions, ultracompact to diffuse.  The full ACIS
field of view is outlined in purple.}
\label{fig:w51mehringer}
\end{figure}

Few examples of true starburst activity in the Milky Way are accessible
to observers, due to large distances and heavy obscuration in the
molecular clouds that give birth to such regions.  W51 is a remarkable
example of such violent, high-mass star formation.  W51A and B are
large star-forming complexes at the tangent point of the Sagittarius
arm, containing dozens of radio HII regions, from hypercompact to
diffuse \citep[][see Figure~\ref{fig:w51mehringer}]{Mehringer94}.
W51C is a very large composite SNR at $D \sim 6$~kpc that may be
physically associated with the W51 complex.  W51 is one of the most
luminous star-forming complexes in the Galaxy; its GMC is in the top
10\% by mass and the top 1\% by size \citep{Carpenter98}.  The many HII
regions contained in the W51 complex probably lie at a range of
distances, from 5.5 to 7.5~kpc.  At $D = 5.5$~kpc, $10^{\prime} \sim
16.0$~pc while for $D = 7.5$~kpc, $10^{\prime} \sim 21.8$~pc.

\citet{Okumura00} estimate the stellar cluster ages in W51A to range
from 0.4 to 2.3 Myr.  The absorption towards these clusters is highly
variable, with $A_V$ ranging from 5 to 40 magnitudes.  Just the main
complex in W51A, called G49.5-0.4, has 35 O stars (the earliest is an
O4); two are supergiants.  The IMF of this region appears to be
top-heavy, with an excess of stars $>30 M_\odot$ and an O-star
formation rate $\sim 25$ times that of the famous HMSFR NGC~3603, called the Galactic analog to 30~Doradus.
W51A may present an example of sequential star formation, where the
winds from earlier epochs of star formation have compressed the
intervening gas to trigger the violent star formation seen in the
central part of G49.5-0.4.

\begin{figure}
\centering
\includegraphics[width=\columnwidth]{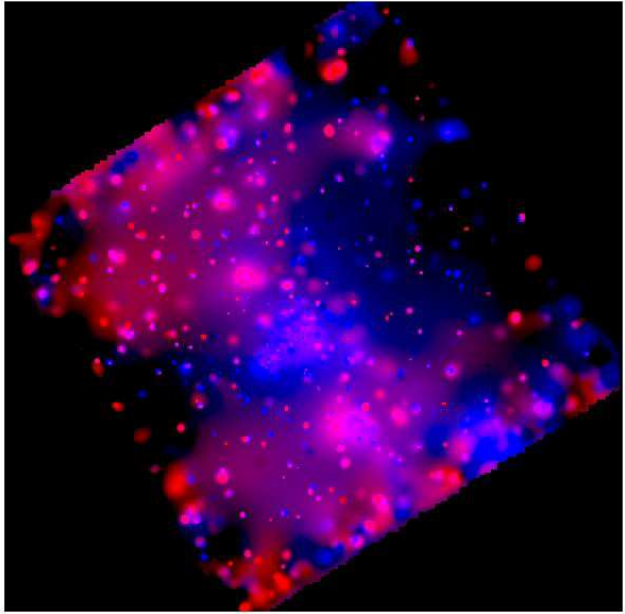}
\caption{\small The 72-ksec ACIS observation of W51A, smoothed with {\it
csmooth}; red is (0.5--2~keV), blue is (2--8~keV).  Only the ACIS-I array
($17^{\prime} \times 17^{\prime}$) is shown.}
\label{fig:w51acisimage}
\end{figure}

A two-color smoothed image of the 72-ksec ACIS-I observation of W51 is
shown in Figure~\ref{fig:w51acisimage}, centered on G49.5-0.4.  Over
450 X-ray point sources are detected in this field, with over 100 of those in
G49.5-0.4.  Diffuse emission is associated with many of the known radio
HII regions.  

\begin{figure}
\centering
\includegraphics[width=\columnwidth]{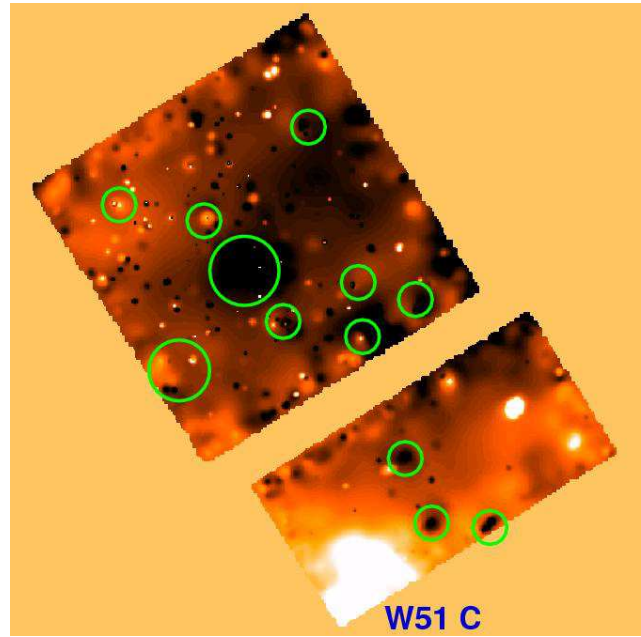}
\caption{\small A hardness-ratio image of W51A, now including the
off-axis CCDs ($8^{\prime} \times 17^{\prime}$) as well as the
ACIS-I array ($17^{\prime} \times 17^{\prime}$); lighter regions
are soft, darker regions are hard.  Some of the radio HII regions in
the field are marked with green circles and the SNR W51C is labeled.}
\label{fig:w51hardness}
\end{figure}

This diffuse emission is perhaps more easily seen in the hardness-ratio
image of the ACIS data, Figure~\ref{fig:w51hardness}.  Faint, hard diffuse
emission may pervade much of the ACIS field.  Several of the
prominent radio HII regions are marked with green circles.  This
observation also serendipitously included the top part of the large SNR
W51C, imaged far off-axis on the backside-illuminated CCD S3.  It is
seen as a soft (white) extended source at the bottom of this image.  A
small part of the W51B star-forming complex is also imaged on the
off-axis CCDs.

\begin{figure}
\centering
\includegraphics[width=\columnwidth]{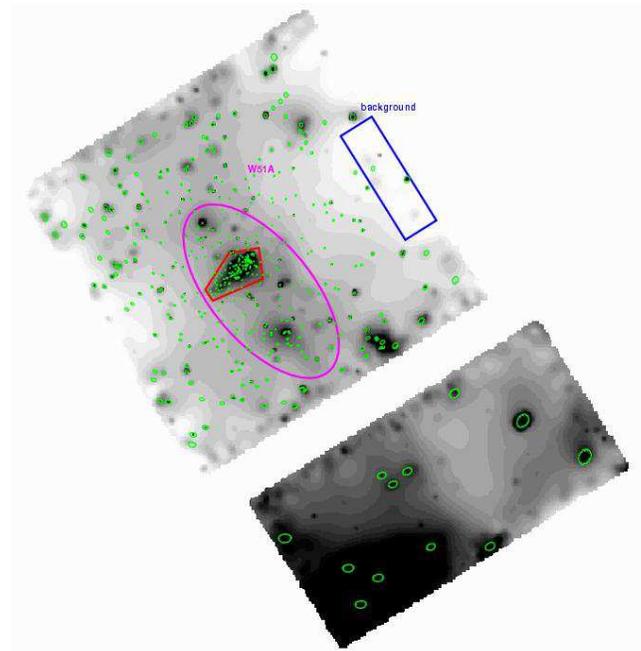}
\caption{\small A full-band (0.5--8 keV) ACIS image of W51, with detected
point sources outlined in green.  Regions used for obtaining the spectrum
of diffuse emission in G49.5-0.4 are described in the text.}
\label{fig:w51fullband}
\end{figure}

As a simple illustration of the spectral properties of the diffuse
emission in W51A, we have extracted and fit the diffuse emission
spectrum from G49.5-0.4.  Figure~\ref{fig:w51fullband} shows the
full-band (0.5--8 keV) ACIS image of W51, with detected point sources
outlined in green.  The magenta ellipse shows the region used to
extract the spectrum of the diffuse emission associated with G49.5-0.4;
point sources were excluded, as was the entire region outlined by the
red polygon, to minimize corruption to the diffuse spectrum from
unresolved young stars in G49.5-0.4.  The region used to estimate the
spectrum of the background is shown by the blue rectangle.  The
background spectrum was scaled to match the geometrical and effective
area of the diffuse emission sample before it was subtracted from the
diffuse spectrum.  We assume a distance of $D = 5.5$~kpc for G49.5-0.4
\citep{Kolpak03}.

\begin{figure}
\centering
\includegraphics[width=0.7\columnwidth,angle=-90]{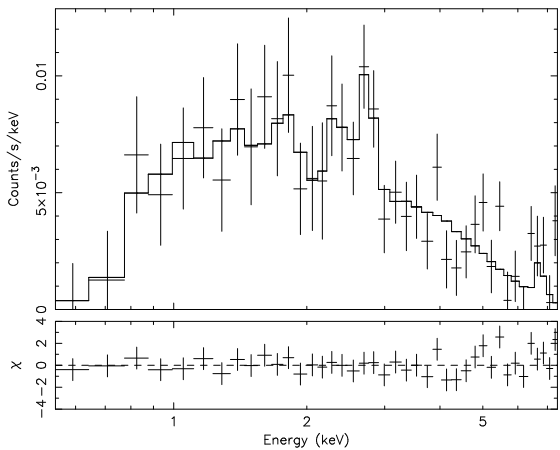}
\caption{\small Diffuse emission from the HII region complex
G49.5-0.4 in W51A.  Two thermal plasma components (both the soft plasma
similar to that seen in M17 and a much harder component) are needed to
fit the spectrum.}
\label{fig:w51spectrum}
\end{figure}

The resulting background-subtracted spectrum and a fit with a
two-temperature thermal plasma model are shown in
Figure~\ref{fig:w51spectrum}.  The soft plasma component ($kT \sim
0.5$~keV) has an absorbing column of $N_H = 1 \times 10^{22}$~cm$^{-2}$
while the hard plasma component ($kT \sim 7$~keV) requires $N_H = 3
\times 10^{22}$~cm$^{-2}$.  This hard component is not well-constrained
by the data but, unlike for RCW49, a power law does not adequately fit
the hard tail in the spectrum.  Two gaussians (at 2.3~keV and 2.7~keV)
were added to achieve an acceptable fit.  As for M17 and RCW49, the
need for these non-physical gaussian components is probably due to a
combination of imperfect background subtraction and averaging over a
large field with many different plasma components and absorbing
columns.  The intrinsic luminosity (absorption-corrected, 0.5--8~keV)
of the $kT \sim 0.5$~keV component is $L_{{\rm X,diffuse}}=1.5 \times
10^{33}$~ergs/s, while that of the $kT \sim 7$~keV component is
$L_{{\rm X,diffuse}}=3.0 \times 10^{33}$~ergs/s.

As with RCW49, this preliminary analysis of one HII region complex in
W51A mainly serves to show that more detailed work is needed to sort out
the various plasma components and absorption characteristics of this
complicated field, although our work is more limited here due to small
numbers of counts.  Although the temperature of the hard plasma component
is not well-constrained by these data, they do confirm that hard diffuse
X-rays pervade HMSFRs when high concentrations of young, early O stars
are present.

\section{Diffuse X-rays from HMSFRs}

{\em {\em Chandra}} has resolved diffuse emission from stellar
populations in several Galactic high-mass star-forming regions. These
are the first unambiguous detections of OB ``windswept bubbles.''
Table~\ref{tab:hmsfrs} (updated from Townsley03) summarizes these
findings, including values from the literature as well as the results
presented here.  The luminosities in this table have been corrected for
absorption.  Diffuse emission is not seen in every HII region --
apparently early and/or multiple O stars (but not necessarily W-R
stars) are required to generate it.  It remains unclear whether this
emission can be generated by individual stars or only via wind-wind
interactions.

\begin{table*}[htb]
\centering
\small
\caption{\small Diffuse X-rays from High Mass Star Forming Regions}\medskip
\label{tab:hmsfrs}
\begin{tabular}{lccccl}
Region & Diffuse       & $N_H$               & kT    & $L_{{\rm X,diffuse}}$ & Reference \\ 
       & Area (pc$^2$) & $10^{21}$~cm$^{-2}$ & (keV) & $10^{33}$~ergs~s$^{-1}$ & \\
\hline\hline
Trifid Nebula		&\ldots &\ldots &  \ldots   &  $< 10^{-2}$ & 
\citet{Rho04}   \\
Orion Nebula            &\ldots &\ldots &  \ldots   &  $< 10^{-3}$ & \citet{Feigelson02}   \\
Eagle Nebula            &\ldots &\ldots &  \ldots   &  $< 10^{-3}$ & \citet{Mytyk01}  \\
Lagoon -- NGC 6530      &\ldots &\ldots &  \ldots   &  $<10^{-2}$  & \citet{Rauw02}  \\
Lagoon -- Hourglass     &  0.04 & 11.1  &    0.63   & $\leq 0.7$   & \citet{Rauw02} \\
Rosette Nebula          &  47   &   2   & 0.06, 0.8 & $\leq 0.6$   & Townsley03 \\
RCW 38                  &   2   &  9.5  & 0.2, $\Gamma=1.6$ & 1.6  & \citet{Wolk02} \\
RCW 49			&  39   & 4, 12 & 0.1, 0.8, 3.1 or $\Gamma=2.3$ & 3.0 & this work \\
Omega Nebula		&  42   &   4   & 0.13, 0.6 &   3.4        & Townsley03 \\
W51A (G49.5-0.4)	&  88   & 10, 28& 0.5, 6.9  &   4.5        & this work  \\
Arches Cluster          &  14   & 100   &    5.7    &    16        &  \citet{YusefZadeh02} \\
NGC 3603                &  50   &   7   &    3.1    &    20        & \citet{Moffat02} \\
Carina Nebula           & 1270  & 3--40 &   0.8:    &   200:       & \citet{Seward82}  \\
\hline
\end{tabular}
\normalsize
\end{table*}   

These results imply that the classical Str{\" o}mgren Sphere model for
HII regions is really better described as a ``Str{\" o}mgren Shell'' in
many cases, at least in those HII regions powered by O stars with
strong winds.  The X-ray flows we see may significantly affect H{\sc
II} region evolution and should be considered in models of HMSFRs.
Such models should accomodate a range of X-ray plasma temperatures ($kT
\sim 0.1-7$~keV) and possibly even non-thermal emission ($\Gamma \sim
2$), a wide range of X-ray luminosities ($L_x \sim 10^{32-34}$~ergs/s),
center-filled morphologies, and diffuse X-ray emission spread over
several-parsec scales.  

The X-ray luminosities that we report are lower limits to the true
emission, due to geometry and obscuration. The Galactic Plane likely
contains substantial soft X-ray emission from HMSFRs, far larger than
what we are able to detect.  X-ray flows from HII regions may
contribute to galaxian features such as the Galactic Ridge emission,
diffuse emission in galaxies, and starburst superwinds.

From Townsley03, we find that only a small portion of the wind energy
and a tiny fraction of the mass appears in the observed diffuse X-ray
plasma in M17.  This energy could be dissipated via turbulence,
mass-loading, fissures in the neutral material surrounding the HII
region, or other processes.  We expect to find similar results for the
other HMSFRs we are studying, based on the preliminary X-ray plasma
temperatures and luminosities presented here.  The hard diffuse X-ray
emission seen in RCW49, W51A, and other HMSFRs containing compact
clusters implies that another, more efficient process is also at work
in regions with many very young, closely spaced OB stars.

This very cursory overview of {\em Chandra} observations of HMSFRs
neglects many interesting details of the data, not the least of which
are the hundreds to $>1000$ X-ray point sources seen in every
observation.  There are many recent and upcoming observations of HMSFRs
with both {\em Chandra} and {\em XMM-Newton}; our table of diffuse X-ray
properties should almost double in length in the next year.  We expect
that our understanding of these fields will be similarly augmented.

\section*{Acknowledgments}

Support provided by NASA contract NAS8-38252 to Gordon Garmire, the
ACIS Principal Investigator.  The 2MASS Atlas images were obtained as
part of the Two Micron All Sky Survey (2MASS), a joint project of the
University of Massachusetts and the Infrared Processing and Analysis
Center/California Institute of Technology, funded by NASA and the
NSF.



\begin{thebibliography}{}
\setlength\itemsep{0cm}




\bibitem[Belloni \& Mereghetti(1994)]{Belloni94} Belloni, T.~\& 
Mereghetti, S.\ 1994, A\&A, 286, 935 

\bibitem[Bruhweiler et al.(2001)]{Bruhweiler01} Bruhweiler, F.,
Bourdin, M., Freire Ferrero, R., \& Gull, T.\ 2001, BAAS, 33, 1450 

\bibitem[Cant\'{o}, Raga, \& Rodr\'{\i}guez(2000)]{Canto00} Cant\'{o}, J.,
Raga, A.~C., \& Rodr\'{\i}guez, L.~F.\ 2000, ApJ, 536, 896

\bibitem[Capriotti \& Kozminski(2001)]{Capriotti01} Capriotti, 
E.~R.~\& Kozminski, J.~F.\ 2001, PASP, 113, 677 

\bibitem[Carpenter \& Sanders(1998)]{Carpenter98} Carpenter, 
J.~M.~\& Sanders, D.~B.\ 1998, AJ, 116, 1856 

\bibitem[Dunne et al.(2003)]{Dunne03} Dunne, B.~C. et al.\ 2003,
ApJ, 590, 306

\bibitem[Feigelson et al.(2002)]{Feigelson02} Feigelson, E.~D. et al.\ 
2002, ApJ, 574, 258

\bibitem[Garmire et al.(2003)]{Garmire03} Garmire, G.~P., Bautz, 
M.~W., Ford, P.~G., Nousek, J.~A., \& Ricker, G.~R.\ 2003, Proc.\ SPIE, 4851, 
28 

\bibitem[Hanson, Howarth, \& Conti(1997)]{Hanson97} Hanson, M.~M.,
Howarth, I.~D., \& Conti, P.~S.\ 1997, ApJ, 489, 698

\bibitem[Kahn \& Breitschwerdt(1990)]{Kahn90} Kahn, F.~D.~\&
Breitschwerdt, D.\ 1990, MNRAS, 242, 209

\bibitem[Kolpak et al.(2003)]{Kolpak03} Kolpak, M.~A., Jackson, 
J.~M., Bania, T.~M., Clemens, D.~P., \& Dickey, J.~M.\ 2003, ApJ, 582, 756 

\bibitem[Mehringer(1994)]{Mehringer94} Mehringer, D.~M.\ 1994, 
ApJS, 91, 713 

\bibitem[Moffat, Shara, \& Potter(1991)]{Moffat91} Moffat, 
A.~F.~J., Shara, M.~M., \& Potter, M.\ 1991, AJ, 102, 642 

\bibitem[Moffat et al.(2002)]{Moffat02} Moffat, A.~F.~J.~et al.\ 2002,
ApJ, 573, 191

\bibitem[Morton(1967)]{Morton67} Morton, D.~C.\ 1967, ApJ, 147, 1017

\bibitem[Mytyk et al.(2001)]{Mytyk01} Mytyk, A.~M.,
Daniel, K.~J., Gagne, M., \& Linsky, J.~L.\ 2001, AAS Meeting, 199, 04.08

\bibitem[Okumura et al.(2000)]{Okumura00} Okumura, S., Mori, A., 
Nishihara, E., Watanabe, E., \& Yamashita, T.\ 2000, ApJ, 543, 799 

\bibitem[Piatti, Bica, \& Claria(1998)]{Piatti98} Piatti, A.~E., 
Bica, E., \& Claria, J.~J.\ 1998, A\&AS, 127, 423 

\bibitem[Pittard, Hartquist, \& Dyson(2001)]{Pittard01} Pittard, J.~M.,
Hartquist, T.~W., \& Dyson, J.~E.\ 2001, A\&A, 373, 1043

\bibitem[Rauw et al.(2002)]{Rauw02} Rauw, G. et al.\ 2002, A\&A, 395, 499

\bibitem[Rho et al.(2004)]{Rho04} Rho, J., Ramirez, S., 
Corcoran, M.~F., Hamaguchi, K., \& Lefloch, B.\ 2004, ArXiv Astrophysics 
e-prints, astro-ph/0401377 

\bibitem[Seward \& Chlebowski(1982)]{Seward82} Seward, F.~D.~\&
Chlebowski, T.\ 1982, ApJ, 256, 530

\bibitem[Townsley et al.(2003)]{Townsley03} Townsley, L.K., 
Feigelson, E.D., Montmerle, T., Broos, P.S., Chu, Y., \& Garmire, G.P.\ 
2003, ApJ, 593, 874

\bibitem[Weaver et al.(1977)]{Weaver77} Weaver, R., McCray, R., Castor,
J., Shapiro, P., \& Moore, R.\ 1977, ApJ, 218, 377

\bibitem[Weisskopf et al.(2002)]{Weisskopf02} Weisskopf, M.~C., 
Brinkman, B., Canizares, C., Garmire, G., Murray, S., \& Van Speybroeck, 
L.~P.\ 2002, PASP, 114, 1 

\bibitem[Whiteoak \& Uchida(1997)]{Whiteoak97} Whiteoak, 
J.~B.~Z.~\& Uchida, K.~I.\ 1997, A\&A, 317, 563 

\bibitem[Wolk et al.(2002)]{Wolk02} Wolk, S.~J., Bourke, T.~L., Smith,
R.~K., Spitzbart, B., \& Alves, J.\ 2002, ApJ, 580, L161

\bibitem[Yusef-Zadeh et al.(2002)]{YusefZadeh02} Yusef-Zadeh, F. et al.\ 2002, ApJ, 570, 665

\end{thebibliography}
\end{document}